\documentclass[slac_one]{revtex4}
\usepackage{graphicx}
\usepackage{fancyhdr}
\begin{document}

\title{New Bound States of several Top-quarks bound by Higgs Exchange}

%

\author{C. D. Froggatt}
\affiliation{Glasgow University}
\author{H. B. Nielsen}
\affiliation{Niels Bohr Institute, Blegdamsvej 17 -21,DK 2100 Copenhagen}

\begin{abstract}
It is suggested that there is a fine tuning principle remarkably
arranging the top-quark Yukawa coupling to be just so as to make
the mass $m_{12}$ of a bound state of 6 top + 6 anti-top quarks become very
small. The value of the top Yukawa needed for Higgs and gluon
etc. exchanges just binding the 12 quarks to make this mass zero is
$g_t|_{m_{12}=0} =1.01 \pm 0.15$, while the experimental running top Yukawa
coupling is $g_t = .94$.
\end{abstract}

\maketitle

\thispagestyle{fancy}


\section{INTRODUCTION} 
Because of the top mass being of the order of e.g. the W-masses,
the top Yukawa coupling is of order unity. The relevant quantity
for estimating when the coupling is sufficiently strong to lead to
non-perturbative effects is rather $\frac{g_t^2/2}{4\pi}$. So it
is only if we could get of the order of $4\pi$ particles to
function together that severe strong interaction effects would
show up, but there are just 12 top quarks and anti-top quarks in
the 1s shell.

\section{HIGGS EXCHANGE}
Like other even order tensor field exchanges - such as the
gravitational field - the scalar Higgs field causes attraction
between all combinations of say top quarks and anti-top quarks. So
it is just the number of tops and anti-tops together that count
for the strength of binding - the binding energy for a bound state
of such quarks. In first approximation, analogous to the Bohr
atom, one finds \cite{FNSurveys} that the binding energy goes as
$E_{binding} \propto (g_t^2Z)^2m_t$, where $Z$ is the number of
particles considered to lie in the centre and attract the
$(Z+1)$'th particle.

For a full 1s shell of $6t + 6\bar{t}$ in the Bohr model
approximation
 we find for $t$- and $u$-channel Higgs exchange:
\begin{equation}
E_{binding} = \frac{11g_t^4}{2\pi^2}m_t
\end{equation}


\section{REMARKABLE COINCIDENCE}
We imagine calculating in the infinite momentum frame so as to make the
in principle relativistic calculation look more non-relativistic.
After including gluon exchange and allowing for the
left-handed top quarks to be replaced by left-handed $b$-quarks, we
arrive at the result that the binding
energy could be so big as to compensate for the mass of the 6 top
and 6 anti-top quarks within the uncertainty in our calculation.
Actually we formulate our calculation so as to evaluate the value
$g_t|_{m_{12}=0}$ of the top Yukawa coupling, which would
correspond to the exact cancellation of the binding energy with
the energy of the masses, so as to make the 12 quark bound state
exactly massless $m_{12}=0$. The resulting value turns out to look
like a remarkable coincidence. We namely included a long series
of corrections in estimating this top Yukawa coupling value and
finally found
\begin{equation}
g_t|_{m_{12}=0} = 1.01\pm 0.15
\label{gtcrit}
\end{equation}
in very good agreement with the experimental value of the
running top Yukawa coupling - which is the one that
should be compared to the value (\ref{gtcrit}) we calculated:
\begin{equation}
g_t^{exp} = .935 \pm .080.
\label{gtexp}
\end{equation}

\section{FINE TUNING CALLS FOR MPP}

This remarkable coincidence presents us with a fine tuning problem
one could say, and thus calls for some principle or law about the
values of coupling constants, such as in the present case the
top-Yukawa coupling.
We have worked for some time on the idea of postulating such a
principle, the so-called ``Multiple Point Principle'' (MPP), which
postulates \cite{MPP}  that there shall be many vacua/phases with
the same energy density.
The approximate masslessness of a particle can in fact be a
consequence of this ``Multiple Point Principle'', when there are
two degenerate phases or vacua just deviating by the particle in
question being condensed in one of the phases while it is not
condensed in the other one.
Such an at least near degeneracy of a condensed and an uncondensed
phase of the bound states we have discussed is anyway unavoidable,
if the mass squared of the bound state is a smooth function
of the coupling constants and in particular of the top Yukawa
coupling in our case. In fact if the coupling is a bit stronger
than our $g_t|_{m_{12}=0}$ it means, in the infinite momentum
frame language, that the bound state becomes a tachyon and that is
well-known to imply that there should be a condensate, i.e. a
non-zero expectation value for an effective field describing the
bound state. On the other hand if the coupling is weaker than our
phase border value $g_t|_{m_{12}=0}$, then the bound state has
positive mass squared and there will be no condensate, i.e. the
expectation value of the field will be zero. The above means that,
taking the coincidence as a fine tuning problem that has to be
explained, then our ``multiple point principle'' at least in this
case is strongly called for.

\section{SPECTRUM OF BOUND STATES}

For bound states of just $Z$ top or anti-top particles, the
binding energy per particle is proportional (approximately) to
$Z^2$ and so
we expect the mass squared $m_Z^2$ of the bound state of $Z$ such
particles to be of the form $m_t^2 Z^2 ( 1- aZ^2)$ where $a$ is a
constant. Assuming, as expected from our infinite momentum frame
thinking, that it is $m_Z^2$ that
shall be analytic, and using that from our fine tuning assumption
the particle with $Z=12$ shall be massless, we get, keeping the
lowest order Taylor expansion terms, that the bound state with $Z$
constituents has the mass squared:
\begin{equation}
m_Z^2 = m_t^2 Z^2(1 -(Z/12)^2).
\end{equation}
Since this formula has the character of an interpolation, we think
it is rather trustable and we obtain for instance for the bound
states with $Z=11$ a mass of $m_{11} = 760$ GeV and for the one
with $Z= 10$ constituents $m_{10} = 950$ GeV.
This means that we expect a
spectrum of related particles of the following form.
\begin{figure}[h]
\centering
\includegraphics[width=70mm]{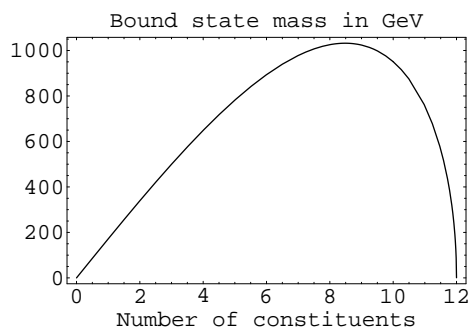}
\caption{Mass spectrum of bound states.} \label{2-f1}
\end{figure}

We can even imagine a $W$ boson being attached to a system of say
$Z=12$ quarks. Then the bound system is in a superposition of being
indeed a system with a $W$ added and one in which, instead, we have
the system that would be obtained by having replaced a left-handed top quark
by a left-handed $b$-quark in the 12-constituent state. We expect such a
bound state with an extra W would have mass closer to that of the
$Z=11$ bound state than to the one with $Z=12$. Either there would be an extra $W$
that would have to be strongly localized - of the order of the
radius - or there would be only 11 top or anti-top quarks although then
together with a $b$-quark too.

\section{PRODUCTION AT LHC OR TEVATRON}
Both the  colour neutral $6t + 6\overline{t}$ and the related
resonances with 11 constituents are strongly bound states and have
small radii $r_0 \sim 1/m_t$ and  their production rate may be
estimated
as if they were ``fundamental particles''
just with some form factor corrections. The bound states with 12
constituents cannot, however, be produced by gluon collision
directly, because they are colour singlets, but can be produced as
decay products from the heavier $6t + 5\overline{t}$ colour triplet
bound states, whose pair production cross section should be similar
to that for a 4th generation quark.
\begin{figure*}[h]
\centering
\includegraphics[width=70mm]{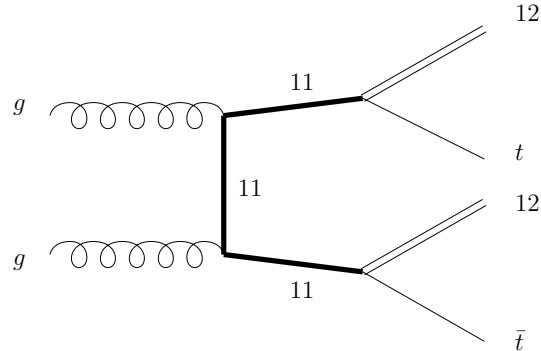}
\caption{Feynman diagram for pair production of the $6t +
5\overline{t}$ colour triplet bound state and subsequent decay.}
\label{feyn_gluongluon-f1}
\end{figure*}

If indeed the 12-constituent bound state is very light it would
decay into a few gluon jets, or if very light might look effectively
like one jet at the high energies needed. This light bound
state would be co-produced with a top quark or anti-top quark resulting
from the decay of the 11-state. These 11-states would only be
produced in pairs and so we should see a $t \bar{t}$ pair being
produced together with two of the light bound states.

\section{OTHER MPP PREDICTIONS}
The existence of a Planck scale degenerate vacuum in the Standard
Model would lead to a Higgs mass prediction:
\begin{center}
$m_h = \left\{ \begin{array} {r@{\quad \quad}l}
132 \pm 5 \ \mbox{GeV} & \mbox{exact degeneracy version} \quad  \mbox{\cite{Mhiggs}} \\
118 \pm 5 \ \mbox{GeV} & \mbox{metastable version} \quad
\mbox{\cite{Meta}}
               \end{array} \right.$
\end{center}
The fine tuning needed in this paper to ensure the existence of a
degenerate phase containing a 12 quark bound state condensate,
together with the existence of a Planck scale vacuum, would solve the
hierarchy problem \cite{FNSurveys} in the Standard Model.
The values of the fine structure constants can be explained using
MPP in the family replicated gauge group model
\cite{FNSurveys,MPP}. The MPP assumption has been adapted to SUGRA
models in order to explain the tiny value of the cosmological
constant \cite{Roman}. The application of MPP to the general 2
Higgs doublet model leads to an approximate custodial symmetry
explaining the absence of flavour changing neutral currents
\cite{Z2}.

\end{document}